\documentclass[twocolumn,amsmath,showpacs,prb,floatfix,superscriptaddress]{revtex4}
\usepackage{graphicx,psfrag,bm,times}
\usepackage{natbib}

\begin{document}

\title{Spontaneous emission interference in negative-refractive-index waveguides}

\author{Gao-xiang Li}
\email{gaox@phy.ccnu.edu.cn}
\affiliation{Max-Planck-Institut f\"ur
Kernphysik, Saupfercheckweg 1, D-69117 Heidelberg, Germany}
\affiliation{Department of Physics, Huazhong Normal University,
Wuhan 430079, China}

\author{J{\"o}rg Evers}
\email{joerg.evers@mpi-hd.mpg.de}
\affiliation{Max-Planck-Institut f\"ur Kernphysik, Saupfercheckweg 1,
D-69117 Heidelberg, Germany}

\author{Christoph H. Keitel}
\affiliation{Max-Planck-Institut f\"ur Kernphysik, Saupfercheckweg 1,
D-69117 Heidelberg, Germany}

\begin{abstract}
The spontaneous decay of a  $V$-type three-level atom placed in a negative-refractive-index waveguide is analyzed. We find that in thin waveguides, highly efficient surface guided modes are supported, which do not occur in positive index waveguides. In addition, at low absorption, the mode density and thus spontaneous emission into particular regular guided modes is enhanced by several orders of magnitude as compared to regular dielectric waveguides.  The asymmetries between emission into the different modes and the enhancement of particular guided modes allow to induce strong spontaneous-emission interference between transitions with orthogonal transition dipole moments.
\end{abstract}

%
%
%
%
%
%
%
%
%
%
%
%
%

\maketitle

\section{Introduction}

Spontaneous emission of atoms is not an immutable property, but can be altered essentially via two different mechanisms. One mechanism involves modification of the internal dynamics. For example, quantum interference among different decay channels of the atom can be established, such that spontaneous emission is modified to a great extend~\cite{s2,s2b}.
A particular class of quantum interference schemes that has received much theoretical attention in the literature is based on so-called spontaneously generated coherences (SGC). These coherences have for example been shown to lead to quenching of spontaneous emission~\cite{s3}, narrow spectral lines~\cite{s4}, phase-dependent line shapes~\cite{s5}, and rapid phase control of collective population dynamics~\cite{s6}.
Despite the large theoretical interest, there is no experimental proof of this type of SGC in atomic systems due to the lack of appropriate candidate systems. The reason for this is that SGC based interference requires the presence of near-degenerate atomic transitions with near (anti-)parallel dipole moments sharing a common atomic state, which does not occur in real atoms~\cite{s2}.
To circumvent this in atoms, schemes to simulate interference or to induce interference by external driving fields have been considered~\cite{s2,s7}, but so far without clear experimental implementation.
An experimental observation of SGC in molecules~\cite{xia} could not be confirmed in a repetition of the experiment~\cite{li}. SGC have been observed, however, in artificial quantum systems, where a suitable level scheme can be designed~\cite{qdot}.
Also, it has been recognized that a more general form of SGC occurs between near-degenerate atomic transitions with near (anti-)parallel dipole moments that do not share a common state. This generalized SGC may lead to measurable effects in realistic atomic systems~\cite{kiffner}. These generalized SGC are also a common interpretation for the occurrence of electromagnetically induced absorption~\cite{eia}.

The second major mechanism for a modification of spontaneous
emission is to modify the  electromagnetic boundary conditions
surrounding the atom, such as in cavities~\cite{s1}. Recently, is
has been shown that a favorable modification of the boundary
conditions is possible in media having  a negative refractive index
(NRI). NRI currently receive a lot of attention because of both
experimental demonstrations and potential applications~\cite{s9}.
For example, NRI material allows to realize superlenses which, in
principle, can achieve arbitrary sub-wavelength
resolution~\cite{s10}. Experimentally, the NRI materials have been
realized over a wide range of frequencies, from the microwave up to
the optical range~\cite{sm, chen}.
It has been shown~\cite{mb} that a single layer of a NRI material
has broadband omnidirectional reflection properties. In the range
between the electric plasma frequency and the magnetic plasma
frequency, the refractive index is close to zero and the NRI
materials reflect radiation for angles of incidence and polarization
with reflectivity of $\sim 0.99$. By exploiting the reflective phase
properties of the NRI materials, an all-dimensional subwavelength
resonator~\cite{hl} has been designed and fabricated.
In the field of quantum optics, two atoms placed at the foci of a
perfect lens formed by a NRI slab exhibit perfect sub- and
superradiance over macroscopic distances~\cite{s11}. Such focussing
and phase compensation can also be used to induce quantum interference~\cite{syz1}.
%
%
\begin{figure}[b]
\centering
\includegraphics[width=\columnwidth]{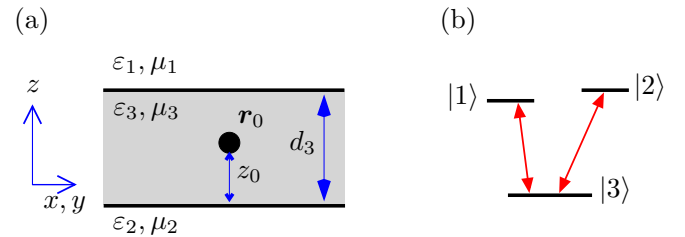}
\caption{\label{fig-1}(Color online) Schematic setup for spontaneous decay in a waveguide
with negative index of refraction material. (a) The waveguide consists of an inner
layer with material parameters $\epsilon_3$, $\mu_3$, and thickness $d_3$. It is surrounded
by infinite outer layers with material parameters $\epsilon_i$, $\mu_i$ ($i\in\{1,2\}$). In the middle layer, an atom is embedded at position $\bf{r}_0$. (b) Electronic structure of the embedded atom, which is a three-level atom in $V$-type configuration.}
\end{figure}
%
%
In a NRI slab waveguide~\cite{s12}, surface guided modes may  exist
for imaginary transverse wave numbers, with power concentrated at
the interfaces rather than inside the slab. In an air waveguide with
NRI cladding~\cite{s13}, both transverse electric and magnetic modes
can be supported with low losses. The existence of the transverse
electric (TE) surface guided modes in the NRI slab waveguide, where
the permeability changes its sign at the interface,  resemble the
transverse magnetic (TM) surface-plasmon-polariton modes at the
interface between the metal and the dielectric, where the
permittivity changes its sign~\cite{bp}. Plasmonic-based
nanophotonic devices have attracted much interest from the quantum
optics community for their use in quantum information
processing~\cite{jl}. A technique which enables strong, coherent
coupling between individual optical emitters and guided plasmon
excitations in conducting nanostructure at optical frequencies has
been proposed~\cite{ml}.  In view of these remarkable properties of
NRI structures, the question arises, how spontaneous emission is
modified in such a surrounding.

Therefore, here we study the spontaneous emission (SE) of an atom
embedded in the middle layer of NRI slab waveguides. The atom is
modelled as a V-type three-level atom with orthogonal dipole moments
on the dipole-allowed transitions as it is the case in real atoms.
The NRI layer is described using a Drude-Lorentz model including
dispersion and absorption. The different contributions to SE are
first classified and interpreted using approximate analytical
results. Then, we verify our results via numerical calculations.
We find that at low absorption, the SE rate into particular
waveguide modes can be several orders of magnitude larger than the
free space rate due to a strong enhancement of the mode density.
Further, in thin waveguides, a dominant contribution to SE arises
from surface guided modes, which do not occur in regular positive
index waveguides. At high absorption, the strong enhancement of
waveguide modes is reduced, while the surface guided mode
contribution remains several orders of magnitude higher than free
space decay.
As our main result, we show that the asymmetric mode structure and the strong enhancement of particular modes can be used to engineer spontaneous emission interference between the two transitions in the embedded atom. Thus we find that the NRI waveguide allows to induce and control near-perfect quantum interference in realistic atomic level schemes.

\section{The model}

We consider a V-type three-level atom embedded in the middle layer of a three-layer waveguide as shown in Fig.~\ref{fig-1}. The upper levels could be Zeeman sublevels \mbox{$|1\rangle = |j\!=\!1,m\!=\!1\rangle$} and \mbox{$|2\rangle= |j\!=\!1,m\!=\!-1\rangle$} with energies $\hbar\omega_1$, $\hbar\omega_2$, and we set the ground state \mbox{$|3\rangle =|j\!=\!0,m\!=\!0\rangle$} energy to zero in the following.
The $y$ direction is the quantization axis (e.g., by applying a weak static magnetic), and the direction normal to the layer interfaces the $z$ axis. Then, the atomic dipole moment operator is given by
\begin{align}
{\bf d}=d(A_{13}{\bf e}_1+A_{23}{\bf e}_2)+H.c.,
\end{align}
where
\begin{align}
{\bf e}_{1,2}=\frac{1}{\sqrt{2}}({\bf e}_z\pm i{\bf e}_x)\,,
\end{align}
and $A_{ij}=|i\rangle\langle j|$ ($i,j\in \{1,2,3\}$) are the atomic transition ($i\neq j$) and
population ($i=j$) operators. ${\bf e}_k$ ($k\in\{x,y,z\}$) are the normalized Cartesian basis vectors, and $d$ is the atomic dipole strength, chosen to be real.
We denote the dielectric permittivity and the magnetic permeability of the layer cladding the atom as $\varepsilon_3(\omega)$ and $\mu_3(\omega)$, which we obtain from a Drude-Lorentz model:
\begin{subequations}
\begin{align}
\varepsilon_3(\omega) &= 1 + \frac{\omega_{pe}^2-\omega_{Te}^2}
 {\omega_{Te}^2-\omega^2-i\omega\gamma_e}\,,
 \\
\mu_3(\omega) &= 1 + \frac{\omega_{pm}^2 - \omega_{Tm}^2}
{\omega_{Tm}^2-\omega^2-i\omega\gamma_m}\,.
\end{align}
\end{subequations}
Here, $\omega_{pe}$ $(\omega_{pm})$, $\omega_{Te}$ $(\omega_{Tm})$ and $\gamma_e$ $(\gamma_m)$ are electric (magnetic) coupling constant, medium oscillation frequency
and linewidth. The permittivity and permeability of the upper [lower] layer, which extends to infinity in the positive [negative] $z$ direction, are $\varepsilon_2(\omega)$ and $\mu_2(\omega)$ [$\varepsilon_1(\omega)$ and $\mu_1(\omega)$].
For these, we assume low absorption and choose $\varepsilon_j$, $\mu_j$ ($j\in\{1,2\}$) as real~\cite{s11,s12,s17}.
$d_3$ is the thickness of the middle layer. The interaction picture Master equation for the density matrix $\rho$ is~\cite{s7}
\begin{align}
\frac{d}{dt}&\rho = \sum\limits_{n=1}^2\Gamma_n(\rho_{nn}A_{33} -
A_{nn}\rho)
\nonumber\\
& + \sqrt{\Gamma_1\Gamma_2}\sum\limits_{n\neq m=1}^2
\kappa_n(A_{mn}\rho-\rho_{nm} A_{33})+\textrm{ H.c.}\,.
\label{eom}
\end{align}
Here, $\Gamma_n$ are spontaneous emission rates~\cite{s14}
\begin{equation}
\Gamma_n= \frac{d^2\omega^2_n}{\hbar\varepsilon_0 c^2} \: {\bf
e_n}^*\cdot {\bf G}_{\rm Im} \cdot {\bf e_n}=\Gamma_{nx}+\Gamma_{nz} \,,
\label{gamma}
\end{equation}
with polarization components $\Gamma_{nx}$ and $\Gamma_{nz}$. The terms involving $\kappa_1$ and $\kappa_2$ in Eq.~(\ref{eom})  are responsible for quantum
interference between the two SE channels $|1\rangle\rightarrow |3\rangle$ and $|2\rangle\rightarrow |3\rangle$, with $\kappa_n$ given by
\begin{equation}
\kappa_n=\frac{d^2\omega_1\omega_2}{\hbar\varepsilon_0
c^2\sqrt{\Gamma_1\Gamma_2}} \: {\bf e_n}\cdot
{\bf G}_{\rm Im} \cdot {\bf
e_n}=\frac{\Gamma_{nz}-\Gamma_{nx}}{\sqrt{\Gamma_1\Gamma_2}}\,.
\label{kappa}
\end{equation}
$-1\leq \kappa_1, \kappa_2 \leq 1$ describes the degree of interference, and ${\bf r}_0=(x_0,y_0,z_0)^T$ is the position of the atom in the middle layer. ${\bf G}_{\rm Im}=\textrm{Im }[{\bf G}({\bf r}_0,{\bf r}_0,\omega_n)]$ is the imaginary part of the electromagnetic Green tensor ${\bf G}({\bf r}_0,{\bf r}_0,\omega)$ given by~\cite{s15}
\begin{align}
{\bf G}({\bf r}_0,&{\bf r}_0,\omega)=  \frac{i\mu_3}{8\pi\tilde{k}_3^2}
\int\limits_0^\infty \frac{dk k}{\beta_3} \left [
({\bf e}_z {\bf e}_z) \:
\frac{2k^2}{D_{p3}}\: {\mathcal I_+^{(p)}}  \right .
\nonumber \allowdisplaybreaks[2] \\
& \left . +
 ({\bf e}_x  {\bf e}_x +{\bf e}_y  {\bf e}_y )
 \left (\frac{\beta_3^2}{D_{p3}} \:
 {\mathcal I_-^{(p)}}  +\frac{\tilde{k}_3^2}{D_{s3}} \:
 {\mathcal I_+^{(s)}}\right ) \right ]\,.
 \label{green}
\end{align}
Here,
\begin{subequations}
\begin{align}
{\mathcal I_\pm^{(q)}} &= (1\pm r^q_{31}e^{2i\beta_3
z_0})(1\pm r^q_{32}e^{2i\beta_3(d_3- z_0)})\,,\\
\tilde{k}_j^2 &=\eta_j\omega^2/c^2\,, \\
\eta_j&=\varepsilon_j(\omega)\mu_j(\omega) \,,
\end{align}
\end{subequations}
with $j\in\{1,2,3\}$. The parameter $k$ is  the magnitude of the vector ${\bf k}=(k_x,k_y)^T$, the conserved component of the wave vector, which is parallel to the interfaces of the layers.  The $\beta_j$ $(j\in \{1,2,3\}) $ are the magnitude of the $z$ component of the wave vector in the $j-th$ layer, whose definition depends on the refraction index of the $j-th$ layer and the value of $k$~\cite{s16}.
If $\textrm{Re}(k)<\textrm{Re}(\tilde{k}_j)$, that is, the corresponding wave in the $j-th$ layer is a propagating one, then $\beta_j$ is expressed as $\beta_j=({\tilde k}_j^2-k^2)^{1/2}$ if the $j-th$ layer has positive refraction index, and $\beta_j=-({\tilde k}_j^2-k^2)^{1/2}$ when the $j-th$ if the layer is a left-handed material. Here, $\textrm{Re}(x)$ is the real part of $x$.
%

\begin{figure}[t]
\centering
\includegraphics[width=8cm]{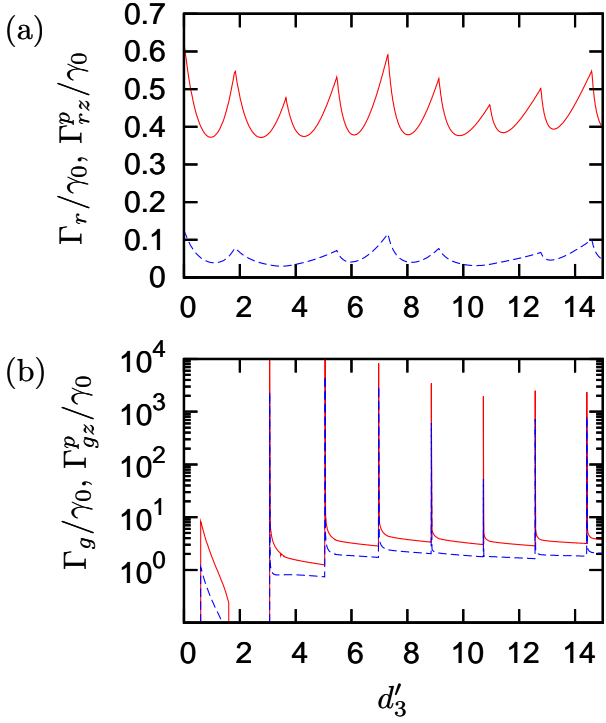}
\caption{\label{fig-2}(Color online) (a) Radiation modes, with
total rate $\Gamma_r$ (red solid line) and contribution $\Gamma^p_{rz}$
of the $p$-mode for $z$-polarization (blue dashed).
(b) Waveguide modes, with
total rate $\Gamma_g$ (red solid line) and contribution $\Gamma^p_{gz}$
of the $p$-mode for $z$-polarization (blue dashed).
The atom is located at $z_0' = 0.25 d_3'$, and the
material parameters are
$\varepsilon_1 = \varepsilon_2 = \mu_1 = \mu_2
= 1.0$, $\omega_{pe}=\omega_{pm}=1.32 \omega_0$,
$\omega_{Te}=\omega_{Tm}= \omega_0$, $\omega_a = 1.09 \omega_0$,
and $\gamma_e = \gamma_m = 10^{-10} \omega_0$. Here, $\omega_0$ is a scaling
parameter. Then,
$\mu_3(\omega_a)=\varepsilon_3(\omega_a) \approx -1.99 + 1.73\cdot 10^{-9}i$.}
\end{figure}

%
%
On the other hand, if $\textrm{Re}(k)>\textrm{Re}(\tilde{k}_j)$ corresponding to an evanescent wave in the $j-th$ layer, then $\beta_j=i(k^2-\tilde{k}_j^2)^{1/2}$ independent of the type of the $j-th$ layer.
Following Ref.~\cite{s15}, we denote the electric field of the TM [TE] wave by the index $p$ [$s$].
The functions $D_{q3}$ ($q\in\{p,s\})$ are defined as
\begin{align}
D_{q3}=1-r^q_{31}r^q_{32}e^{2i\beta_3 d_3}\,,
\end{align}
where $r^q_{31}$ and $r^q_{32}$ are reflection coefficients given by
\begin{equation}
r^p_{ij}=\frac{\varepsilon_j
\beta_i-\varepsilon_i\beta_j}{\varepsilon_j
\beta_i+\varepsilon_i\beta_j}, \qquad r^s_{ij}=\frac{\mu_j
\beta_i-\mu_i\beta_j}{\mu_j \beta_i+\mu_i\beta_j} \,.
\end{equation}
From Eq.~(\ref{green}) we can see that the inhomogeneity of the medium along the $z$ axis leads to a spatially asymmetric Green tensor, as the $z$-component is different from those in the $x-y$ plane.
For notational simplicity, in the following $\beta_3$ and $k$ will be re-scaled by $\omega/c$, i.e, $\beta_3/(\omega/c)$ and $k/(\omega/c)$ will be replaced by $\beta_3$ and $k$. As usual, the permittivity and permeability coefficients of the three layers are assumed to obey $\textrm{Re}(\eta_3) > \textrm{Re}(\eta_1) \ge \textrm{Re}(\eta_2)$.
Then the electromagnetic modes in this structure can be classified into radiation modes with $0<Re(k^2)<Re(\eta_2)$, substrate modes with $Re(\eta_2) <Re(k^2)<Re(\eta_1)$, regular guided modes with $Re(\eta_1)<Re(k^2)<Re(\eta_3)$, and evanescent modes with $Re(k^2)>Re(\eta_3)$.  As quantum interference requires near-degenerate transition frequencies, we also assume $\omega_1\approx\omega_2=\omega$, so that $\Gamma_1\approx \Gamma_2=\Gamma$ and $\kappa_1\approx\kappa_2=\kappa$.

\section{Results}

We assume that both the upper and the lower layers are dielectric media with positive refraction index, and the middle layer surrounding the atom is left-handed material. This structure is a so-called negative-refraction-index (NRI) waveguide~\cite{s12}.
The contributions of the radiation and the substrate modes to the SE rate can be obtained by integrating Eq.~(\ref{green}) numerically. An example is shown in Fig.~\ref{fig-2}(a).
For the following analytical considerations, we assume $\varepsilon_3$ and $\mu_3$ to be constant and real, i.e., we neglect absorption and dispersion. But our numerical results include both absorption and dispersion.

\subsection{Waveguide modes}

\begin{figure}[t]
\begin{center}
\includegraphics[width=8.5cm]{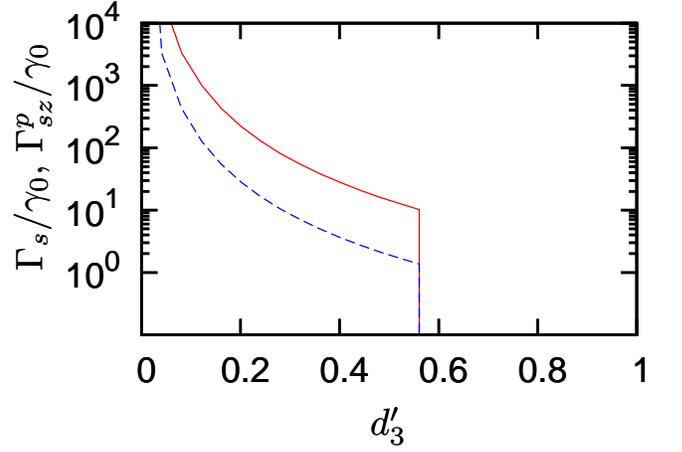}
\end{center}
\caption{\label{fig-3}(Color online) Decay rates into surface guided
modes, with total decay rate contribution $\Gamma_s$ and
contribution $\Gamma^p_{sz}$ of the $p$-mode for $z$-polarization.
The parameters are as in Fig.~\ref{fig-2}.}
\end{figure}
%
The regular guided modes have complex reflection coefficients with modulus 1. In this region, the imaginary parts of the integrands in Eq.~(\ref{green}) are zero, apart from resonances when $D_{q3}=0$ $(q\in\{p,s\})$ is satisfied.
For example, the SE rate of the $z-$ component of the atomic dipole moment into $p-$polarized guided modes is
\begin{align}
\Gamma^p_{gz}&= \frac{3\pi\gamma_0}{4|\varepsilon_3|}
\left. \sum_{m} \frac{k^2\{1+\cos[2(\beta_3
z_0^\prime+\phi^p_{32})]\}}
{\left |d_3^\prime+\frac{\varepsilon_1\varepsilon_3}
{\beta_1}\,\chi_1 +
\frac{\varepsilon_2\varepsilon_3}{\beta_2}\,\chi_2 \right |} \right |_{k=k^{(m)}_{gp}},
\label{gamma-gz}
\end{align}
in which $z_0^\prime=z_0\omega/c$, $d_3^\prime =d_3\omega/c$, and $\gamma_0$ represents the SE rate of the atom in free space. Further,
\begin{align}
r^p_{32}=\exp(-2i\phi^p_{32})
\end{align}
with $0\leq \phi^p_{32}\leq \pi/2$, and
\begin{align}
\chi_i =
\frac{\varepsilon_3\mu_3-\varepsilon_i\mu_i}{\varepsilon_i^2\beta_3^{
2}+\varepsilon_3^2\beta_i^2}\,.
\end{align}
The parameters $k_{gq}^{(m)}$ $(m\in \{1,2,\dots,m^p_{max}\})$ represent the wave numbers of the $m$th $p-$polarized guided modes in the $x-y$ plane, which  are the real roots of equation $D_{p3}=0$ within the region of $\sqrt{\eta_1} <k<\sqrt{\eta_3}$.
The number of modes is $m^p_{max}$, which depends on the thickness of the middle layer, the mode polarization, and the material parameters.
Evidently, each guided mode corresponds to a standing wave in this structure, and the SE rate depends on the position of the atom. With increasing thickness $d_3$, the contribution of the guided modes to the SE rate exhibits a sharp cusp at the
appearance of an extra mode. For the symmetric case $\varepsilon_1=\varepsilon_2$, $\mu_1=\mu_2$ and $z_0=d_3/2$, there is no contribution to the decay rate
from the $p-$polarized modes whose nodes are exactly coincident with the atomic location, such that $\beta_3 d_3+2\phi_{32}^p = (2n+1)\pi$ $(n\in \{0,1,2,\dots \})$. Only modes with an anti-node coinciding with the atomic position contribute to the decay rate, since then $\beta_3 d_3+2\phi_{32}^p=2n\pi$.

Full numerical results for the guided modes are shown in Fig.~\ref{fig-2}(b).
Note that guided modes may already occur for very thin, even sub-wavelength, layers, as the Goos-H\"{a}nchen phase shifts (GHPS) at the interfaces between negative and positive layers enhance the phase change induced by the optical path~\cite{s17}.
The second and the third terms in the denominator of Eq.~(\ref{gamma-gz}) arising from the GHPS are negative, while the first term due to the optical path is positive. Via this cancelling, in $z$-direction the structure effectively acts as a planar cavity with tiny length and large mode density.
In Fig.~\ref{fig-2}, the decay rate is enhanced by about $4$ orders of magnitude compared to the free space decay rate for specific waveguide geometries.
This is in sharp contrast to the regular dielectric waveguide with $\varepsilon_3,\mu_3>0$, where all three terms in the denominator are positive, such that the guided modes have only a small density of modes.
Thus in a suitable NRI waveguide, the atom spontaneously emits photons into few guided modes with large amplitudes. For an atom placed close to a nano-structure,
spontaneous emission can be greatly enhanced due to couplings to electronic quasi-particle surface excitations~\cite{s18}. But here, the enhancement of some guided
modes arises from a variation of the summation of the optical path and the phase changes at the interfaces between negative and positive refraction index materials. Thus these interfaces lead to the strong enhancement.

\subsection{Surface guided modes}
In regular dielectric waveguides, evanescent modes do not
contribute to the SE rate, as they
cannot propagate out.
But if one or
two of the layers are made from NRI, surface
guided modes with $k^2>\eta_3$ can exist, whose
wave vectors obey the resonance
condition $D_{q3}=0$  in the Green's function
Eq.~(\ref{green}).
%
%
\begin{figure}[t]
\begin{center}
\includegraphics[width=8.5cm]{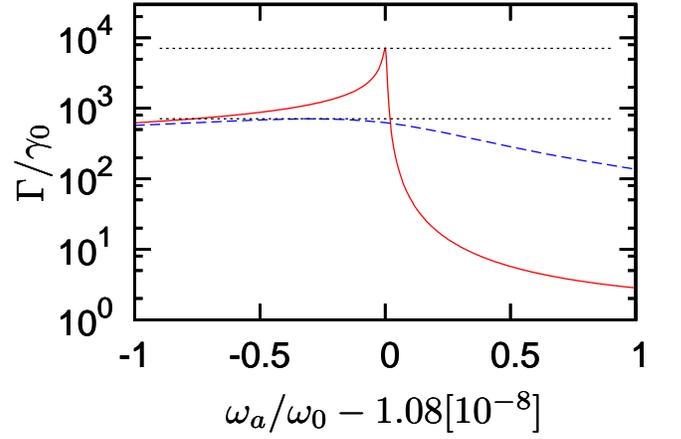}
\end{center}
\caption{\label{fig-4}(Color online) The influence of dispersion and absorption
on the results. The red solid line
shows the frequency dependence of the total decay rate for parameters
as in Fig.~\ref{fig-2}, but with $d_3$ set to the first peak in
$\Gamma_z^p$. The blue dashed line is plotted
for same parameters but includes small absorption $\gamma=\gamma_e=\gamma_m=10^{-8}\omega_0$.
The black horizontal dotted lines indicate the peak amplitudes of
about $7100\gamma_0$ and $710\gamma_0$, respectively, of the two curves,
verifying the scaling of the peak amplitudes with $1/\sqrt{\gamma}$.}
\end{figure}
%
Different from the regular positive index dielectric waveguide, where the
magnitudes of the reflection coefficients $r^q_{32}$ and $r^q_{31}$
are smaller than 1, in the NRI waveguide,
the modulus of $r^q_{32}$ and $r^q_{31}$ can be larger than 1.
Thus real roots $k^{(m)}_{sq}$ $(m\in \{1,2,\dots\})$ can
be found, corresponding to the wave vectors of the surface guided modes.
The contribution of these guided modes to the
$z-$component of the dipole moment is
\begin{eqnarray}
\Gamma^p_{sz}&=&\left.  \frac{3\pi\gamma_0}{4|\varepsilon_3|}
\sum_{m}\frac{k^2\{\cosh[2(\beta_3 z_0^\prime-\phi^{sp}_{32})]-1\}}
{\left |d_3^\prime-\frac{\varepsilon_1\varepsilon_3}{\beta_1}\chi_1
-\frac{\varepsilon_2\varepsilon_3}{\beta_2}\chi_2 \right |}
\right |_{k=k^{(m)}_{sp}} \,.
\label{gamma-sz}
\end{eqnarray}
Here we have set $r^{p}_{32}=-\exp(2\phi^{sp}_{32})$ with $\phi^{sp}_{32}\geq 0$. In Eq.~(\ref{gamma-sz}), the mode functions are hyperbolic instead of the standing wave functions for the regular guided modes Eq.~(\ref{gamma-gz}).

\begin{figure}[t]
\centering
\includegraphics[width=8cm]{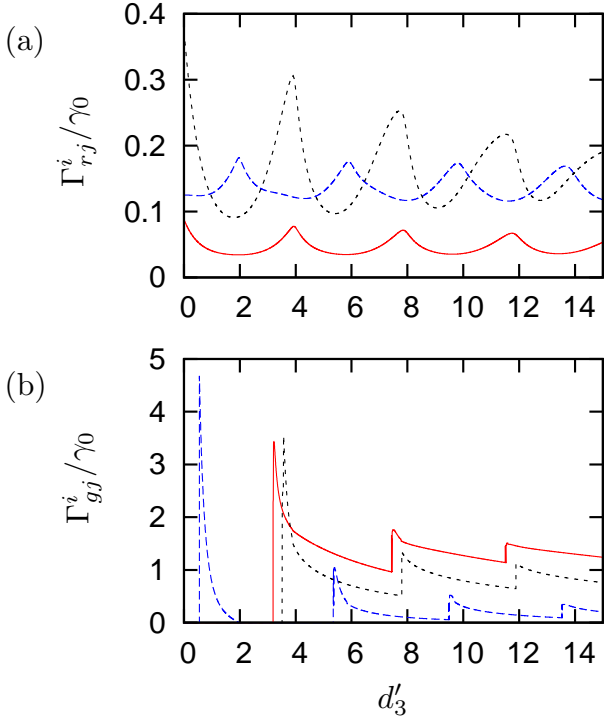}
\caption{\label{fig-5}(Color online) Spontaneous emission
into strongly absorbing NRI waveguides. (a) Radiation modes, with
total rate $\Gamma_r$ (red solid line) and contribution $\Gamma^p_{rz}$
of the $p$-mode for $z$-polarization (blue dashed).
(b) Waveguide modes, with
total rate $\Gamma_g$ (red solid line) and contribution $\Gamma^p_{gz}$
of the $p$-mode for $z$-polarization (blue dashed).
The atom is located at $z_0' = 0.5 d_3'$, and the
material parameters are
$\varepsilon_1 = \varepsilon_2 = \mu_1 = \mu_2 = 1.0$,
$\omega_{pe}=1.25 \omega_0$, $\omega_{pm}=1.189 \omega_0$,
$\omega_{Te}=\omega_{Tm}= \omega_0$,
$\omega_a = 1.08 \omega_0$,
and $\gamma_e = \gamma_m = 10^{-3} \omega_0$.
Here, $\omega_0$ is a scaling parameter. Then,
$\varepsilon_3(\omega_a) \approx -2.38 + 2.19\cdot 10^{-2}i$ and
$\mu_3(\omega_a) \approx -1.48 + 1.61\cdot 10^{-2}i$.}
\end{figure}

By numerical and analytical inspection,
we find that there are two kinds of surface guided modes. The first
has a wave vector $k^2$ close to $\eta_3$,
while the other one has $k^2 \gg \eta_3$.
%
It can be proven that the surface guided
modes with $k^2\gg \eta_3$ can only exist when the
thickness $d_3$ of the middle layer is very thin. As an example,
we estimate the condition for this kind of surface
guided modes for the symmetric NRI
waveguide ($\varepsilon_1=\varepsilon_2$, $\mu_1=\mu_2$).
For the $p-$polarized surface guided modes, the
condition can be approximated as
\begin{align}
k^{(m)}_{sp}d_3^\prime = \ln \left |\frac{\varepsilon_1-\varepsilon_3}{\varepsilon_1+\varepsilon_3} \right |\,.
\end{align}
Therefore $p-$polarized surface modes with $k^2\gg \varepsilon_3\mu_3$ may exist only if the thickness $d_3$ obeys $d_3^\prime\ll \ln|(\varepsilon_1-\varepsilon_3)/(\varepsilon_1+\varepsilon_3)|/ \eta_3$, i.e., $d_3$ is much smaller than one wavelength. For these surface modes,
Eq.~(\ref{gamma-sz}) reduces to
\begin{equation}
\Gamma^p_{sz}= \left .\frac{3\pi\gamma_0}{4|\varepsilon_3|}
\sum_{m}\frac{k^2\{\cosh[\beta_3(d_3^\prime-2
z_0^\prime)]-1\}}
{\left |d_3^\prime-\frac{2\varepsilon_1\varepsilon_3
(\varepsilon_3\mu_3-\varepsilon_1\mu_1)}
{k^3 (\varepsilon_1^2-\varepsilon_3^2)}\right |} \right |_{k=k^{(m)}_{sp}}.
\label{type-1}
\end{equation}
We can see that if $z_0^\prime\neq d_3^\prime/2$, then the decay
rate  can be large, as shown in Fig.~\ref{fig-3}, as this mode only
survives for thin $d_3$ with a large amplitude. This large
enhancement of surface guided mode excitation enables dipole
emission to be preferentially coupled to the surface guided modes,
which may be applied in creating well-guided light sources at the
nanoscale.

Regarding the second type of surface guided modes with $k^2$ close to $\eta_3$, it can be found that these modes may exist if
\begin{align}
d_3^\prime \leq d_3^{max} = - \frac{2\varepsilon_1}{\varepsilon_3 \sqrt{\varepsilon_3\mu_3 -\varepsilon_1\mu_1}}\,.
\end{align}
This is the reason for the sharp cutoff visible for the solid curves in Fig.~\ref{fig-3}. The decay rate induced by these $p-$polarized surface guided modes can be approximated as
\begin{equation}
\Gamma^p_{sz}=\frac{3\pi\gamma_0|\mu_3|}{8}
\sum\limits_{m}\frac{\left [(k^{(m)}_{sp})^2
-\varepsilon_3\mu_3 \right ]
(d_3^\prime-2 z_0^\prime)^2}
{\left |d_3^\prime- d_3^{max} \right |}.
\label{gpsz}
\end{equation}
In thin waveguides, these contributions are small as compared to
those of Eq.~(\ref{type-1}), due to the difference of the mode
amplitudes.

Thus for thin NRI waveguides, the atom can emit strong evanescent fields with large wave number $k^2\gg \eta_3\mu_3$ in the $x-y$ plane. This result is similar to experimental evidence employing a pristine silver film with natural roughness as a NRI slab~\cite{s20}, where the transmission of evanescent waves rapidly grows with the NRI film thickness up to a thickness of about $\lambda/10$.


\begin{figure}[t]
\begin{center}
\includegraphics[width=8.5cm]{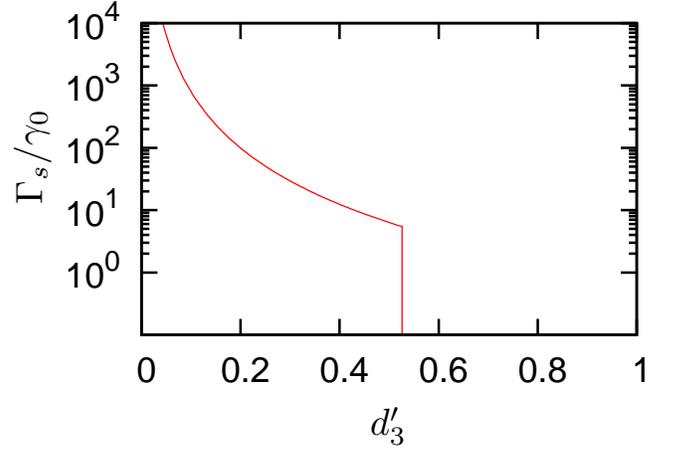}
\end{center}
\caption{\label{fig-6}(Color online) Decay rates into surface guided
modes. The total decay rate $\Gamma_s$ has contributions only from
$\Gamma^p_{sx}$ of the $p$-mode for $x$-polarization. The parameters
are as in Fig.~\ref{fig-5}.}
\end{figure}
%

\subsection{Dispersion and absorption}

With absorption and dispersion, the above equations are more complex. Then, the sharp peaks due to the guided modes become Lorentzians, with widths depending on the imaginary part of the refraction index. For small $\gamma=\gamma_e=\gamma_m$, their amplitudes are proportional to $1/\sqrt{\gamma}$. This scaling generalizes our  results to moderate absorption strengths. An example is shown in Fig.~\ref{fig-4}. This figure shows a magnification on the first peak at $d_3^\prime \approx 3$ in the waveguide decay $\Gamma^{p}_{gz}$ shown in Fig.~\ref{fig-2}. It can be seen that with increasing absorption, the narrow peak is broadened such that the peak amplitude is reduced, in accordance with our scaling law. In contrast to Fig.~\ref{fig-2}, in Fig.~\ref{fig-4} the frequency-dependence of the results is shown in order to analyze the dispersive
properties of the NRI waveguide.

We now turn to the case of high absorption. An example for decay into radiation modes in a strongly absorbing waveguide is shown in Fig.~\ref{fig-5}(a). The corresponding results for waveguide modes and special guided modes are shown in Fig.~\ref{fig-5}(b) and Fig.~\ref{fig-6}. It can be seen that while the waveguide modes are strongly reduced to peak decay rates of order $\gamma_0$, the radiation modes and special guided modes are much less affected by the absorption. Speaking pictorially, the absorption leads to a washing out of narrow structures found in the waveguide modes, such that their amplitude is strongly reduced with increasing absorption. In contrast, the structures in the radiation and special guided modes are already rather broad at low absorption such that they do not change much towards higher absorption.

\subsection{Quantum interference}

\begin{figure}[t]
\centering
\includegraphics[width=8.5cm]{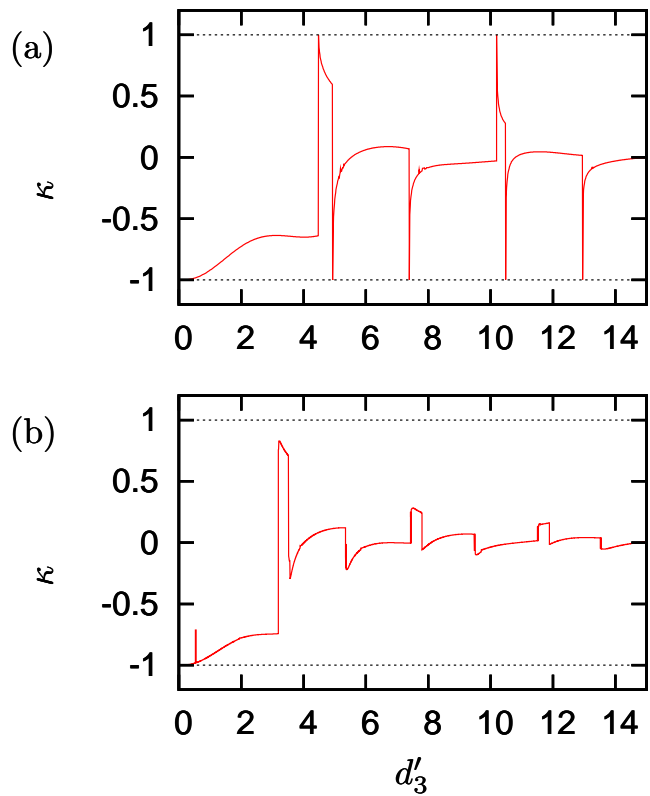}
\caption{\label{fig-7}(Color online) Quantum interference induced by a
negative-refraction-index waveguide structure, characterized
by the interference strength $\kappa$.
(a) Parameters as in Fig.~\ref{fig-2}
except for $z_0^\prime = d_3^\prime/2$, $\omega_{pm}=1.189 \omega_0$
[$\mu_3(\omega_a)\approx -1.20 + 1.27\cdot 10^{-9}i$].
(b) Strong absorption case. Parameters are as in Fig.~\ref{fig-5},
with
$\varepsilon_3(\omega_a) \approx -2.38 + 2.19\cdot 10^{-2}i$
and $\mu_3(\omega_a) \approx -1.48 + 1.61\cdot 10^{-2}i$.}
\end{figure}

Finally, we evaluate the cross-coupling terms $\kappa_n$ in Eq.~(\ref{eom}), which are responsible for SE interference. The strong enhancement of decay into
particular guided modes at low absorption can be used to generate controllable, near-perfect quantum interference, i.e. $\kappa \approx \pm 1$, as shown in
Fig.~\ref{fig-7}(a).
If both transitions mainly interact with the same strongly enhanced guided mode, then they couple even though they have orthogonal transition dipole moments.
For example, if $d_3^\prime$ is chosen close to the birth of a $s-$polarized mode, but away from $p$-polarized mode maxima, then the decay rate is dominated by $\Gamma_{x}^s$ induced by the $s-$polarized modes to the $x$-component of the atomic dipole.
It follows from Eq.~(\ref{kappa}) that then $\kappa\approx -1$, i.e., strong quantum interference similar to that of a hypothetical atom having two near-degenerate transitions with near-antiparallel dipole matrix elements in free space~\cite{s2,s3,s4,s5,s6}.
A second example is for atoms located at the center of the NRI layer, with $d_3^\prime$ set at the birth of a  $p$-polarized mode. If $\beta_3d_3^\prime+2\phi_{j}^p=(2n+1)\pi$, then the decay rate is dominated by $\Gamma_{gx}^p$ from the $x-$component of the atomic dipole, as the atom is at the node of the $z$-modes. But for $\beta_3d_3^\prime+2\phi_{j}^p=2n\pi$, it is dominated by $\Gamma_{gz}^p$, as the atom is at the node of the $x$-modes.
These situations yield strong quantum interference with $\kappa\approx \pm 1$, similar to near (anti-) parallel dipole moments in free space.
%
%
%
Both cases are shown in Fig.~\ref{fig-7}(a).
The extremal values of $\kappa$ occur at
peaks in the SE
rate due to the enhanced mode density, and thus $d_3^\prime$
controls $\kappa$ between $-1$ and 1.
In Fig.~\ref{fig-7}(a),
we have used $\varepsilon_3\neq \mu_3$
to enable the first mechanisms for extremal $\kappa$, and
$z_0^\prime=d_3^\prime/2$ to allow for the second mechanism.
Thus NRI waveguide structures allow to effectively
induce spontaneous-emission quantum interference with realistic
atomic level structures.

Interestingly, spontaneous emission interference can already
be obtained at high absorption, even though in this case the
narrow peaks of large amplitude are absent from the waveguide
mode spectrum. An example is shown in Fig.~\ref{fig-7}(b) for
the parameters of Fig.~\ref{fig-5}. In particular at small
thicknesses $d_3^\prime$ of the waveguide, a large degree
of interference is achieved ($\kappa\approx-1$). The reason for
this is the dominant contribution of the special guided modes.
For the particular setup chosen in this figure, only special
waveguides in the $p-x$ mode are excited. But also at
larger thicknesses, interference of alternating sign in $\kappa$
is achieved, even though with $|\kappa|$ smaller than unity.
This has to be compared, however, to a value of $\kappa=0$ for
atoms in free space.

\section{Summary}

In conclusion, we have investigated the spontaneous decay of a
three-level $V$-type atom placed in a three-layer waveguide
with negative-refraction-index material as middle layer.
We have found that spontaneous emission into particular guided modes
can be greatly enhanced in materials with low losses.
Due to a large mode density, spontaneous decay rates into these modes can be increased by several orders of magnitude as compared to regular dielectric waveguides.
Both at low and high absorption, NRI waveguides support additional
surface guided modes, which do not occur in positive index
waveguides. The modes also offer a strong enhancement of spontaneous
decay compared to the free space case, and are especially effective in thin waveguides.
The specific properties of the waveguide as well as the position of the atom in the waveguide enable one to modify the emission into the different modes to a great extend. We have shown that this control can be used to achieve asymmetries in the decay into modes with different polarizations. This feature allows to induce strong spontaneous emission interference in realistic atomic level schemes, which do not exhibit such interference in free space.

In addition to NRI waveguides, we also analyzed air-like waveguides with NRI cladding~\cite{s13}. In these structures, the middle layer is air-like, while the upper and the lower layer are made of NRI material. We obtained similar results as for the NRI waveguides. In particular, also in this case the density of the guided modes and thus SE can be very large if the absorption is low.

\begin{acknowledgements}
Gao-xiang Li gratefully acknowledges financial support from the
Alexander von Humboldt Foundation, the National Natural Science
Foundation of China (under grant Nos. 10674052 and 80878004), and
the Ministry of Education under project NCET (under Grant No
NCET-06-0671). JE gratefully acknowledges hospitality during his
stay in Huazhong Normal University. Helpful discussions with 
S. Y. Zhu are gratefully acknowledged.
\end{acknowledgements}

\end{document}